# Accelerating Motif Finding in DNA Sequences with Multicore CPUs

Pramitha Perera and Roshan Ragel, *Member, IEEE*

*Abstract* – Motif discovery in DNA sequences is a challenging task in molecular biology. In computational motif discovery, Planted (l, d) motif finding is a widely studied problem and numerous algorithms are available to solve it. Both hardware and software accelerators have been introduced to accelerate the motif finding algorithms. However, the use of hardware accelerators such as FPGAs needs hardware specialists to design such systems. Software based acceleration methods on the other hand are easier to implement than hardware acceleration techniques. Grid computing is one where such software based acceleration technique has been used in acceleration of motif finding. However, drawbacks such as network communication delays and the need of fast interconnection between nodes in the grid can limit its usage and scalability. As using multicore CPUs to accelerate CPU intensive tasks are becoming increasingly popular and common nowadays, we can employ it to accelerate motif finding and it can be a faster method than grid based acceleration. In this paper, we have explored the use of multicore CPUs to accelerate motif finding. We have accelerated the Skip-Brute Force algorithm on multicore CPUs parallelizing it using the POSIX thread library. Our method yielded an average speed up of 34x on a 32-core processor compared to a speed up of 21x on a grid based implementation of 32 nodes.

## I. INTRODUCTION

Deoxyribo Nucleic Acid (DNA) sequences contain genes that code genetic information in living beings. Motifs are short nucleotide patterns locate near these genes. They occur repeatedly in the sequence with mutations in some of their nucleotide positions. DNA motifs often represent Transcription Factor Binding Sites (TFBS) where proteins such as Transcription Factors bind to these sites to regulate the expression of genes. Hence discovery of such motifs helps to understand the mechanisms of gene expression [3].

In the past, motif discovery was carried out using typical methods such as DNase foot printing and gel-shifts [1]. Computational motif discovery has emerged with the advancement of computational biology and become an extensively studied area of research due to its importance. In order to assess computational algorithms for motif discovery many versions of motif finding problem have been formulated. Planted (l, d) motif finding problem (or Planted Motif Search - PMS) is a widely addressed problem in many literature. It is a simplified combinatorial problem of biological motif discovery introduced by Pevzner and Sze [5].

Pramitha Perera is with the Department of Computer Science and Statistics, University of Peradeniya, Sri Lanka (email: pramithap@gmail.com).

Roshan Ragel is with the Department of Computer Engineering, University of Peradeniya, Sri Lanka (email: roshanr@pdn.ac.lk).

### A. Planted (l, d) Motif Finding Problem

Planted (l,d) motif finding problem can be described as follows: if there are $n$ number of sequences of length $t$, and each sequence is planted with an instance of the consensus motif $m$ of length $l$ each having at most $d$ mutations, determine the consensus motif.

Motif consensus is the motif that does not have any mutations. Rest of the motif instances emerge from the consensus motif. There are more than hundred algorithms available to solve this problem [2], [3]. Such algorithms can either be exact or approximate. Exact algorithms such as Brute Force, Planted Motif Search (PMS), CONSENSUS which are based on exhaustive enumeration always identify the correct exact motif whereas approximate algorithms such as Gibbs Sampling, MEME which are based on probabilistic models sometimes fail to identify the correct motif. However, many algorithms take a very long time to solve the challenge problem proposed by Pevzner and Sze [4].

### B. The Challenge Problem

The challenge problem is defined as follows [4]: Find the motif in a set of random DNA sequences which are 600 nucleotides long. Each sequence is implanted with one motif instance which is of length 15 with 4 mutations. This is called the (15, 4) motif problem. In the (15, 4) Fixed number of Mutations model (FM), each motif instance is made by mutating 4 random positions of a 15 nucleotides long motif consensus.

### C. Problem Statement and Proposed Solution

To accelerate motif finding, numerous hardware based acceleration mechanisms such as using Field Programmable Gate Arrays (FPGA) and Graphics Processing Units (GPU) have been introduced. More often to develop FPGA systems special designers are needed and special hardware is needed for GPU based acceleration. Therefore, software based acceleration techniques which do not require such designs and easier to implement than hardware based methods have been introduced. One such software based acceleration that has been introduced is Grid Computing [6]. However, the grid based acceleration suffers from limitations such as network communication delays between nodes which can limit the usage of such methods.

Nowadays multicore CPUs are being used to accelerate many applications that require large amount of data processin such as string matching applications and has gained significant throughput [8]. Therefore, the use of multicore CPUs to accelerate motif finding problem can be advantageous and should achieve better performance over the use of a grid network.

In order to prove this hypothesis, we implemented the enhanced brute force algorithm using the POSIX thread library to run on multicore CPUs. The enhanced brute force algorithm is an exact motif finding algorithm and was used in [6] for their grid based acceleration. We were able to obtain 34x average speed up over single thread implementation on a 32 cores CPU that supports 64 threads, much better than the 21x speedup obtained in [6].

The rest of the paper is organized as follows: In Section II related works are discussed and in Section III background details are discussed. Sections IV and V present experimental methodology and results obtained respectively. Finally in Section VI, we conclude the paper.

## II. RELATED WORK

Many mechanisms have been proposed to accelerate both exact and approximate algorithms for motif finding. In this section, we briefly describe methods that have been proposed to accelerate different motif finding algorithms. Main acceleration techniques discussed in research include the use of FPGAs, GPUs and grid computing.

The Multiple Expectation Maximization for Motif Elicitation (MEME) is a popular and efficient approximate algorithm used by researchers. Chen et al. [10] have accelerated this algorithm using GPUs and have achieved significant speed up over its parallel version. The authors claim that more speed up can be achieved by using a cluster of GPUs. Liu et al. [11], [12] have extended this work by accelerating it in Compute Unified Device Architecture (CUDA) enabled GPUs and also using multiple GPUs. They have compared its performance with the parallel MEME running in CPU clusters and have shown that GPU based acceleration is much better than use of CPU clusters.

Exact motif finding algorithms have also been used in research. Farouk et al. [7] have implemented an enhanced version of the brute force motif finding algorithm known as Skip-Brute Force search on FPGA. On FPGA matching units have been designed in hardware to achieve parallelism. Their implementation has achieved significant speed up over the serial version. The algorithm we took to perform our multicore based acceleration is also the enhanced brute force search algorithm used by Farouk et al.

Grid computing is an emerging architecture for accelerating many algorithms that needs high amount of data processing. With that trend, in a paper published in 2010 [7], Faheem has parallelized the enhanced brute force algorithm for motif finding to run in EUMEDGRID structure. This is a software based acceleration method which has used Message Passing Interface (MPI) for parallel programming. He has tested the program with varying number of worker nodes on the grid. The grid computing based implementation has gained significant speedup over the sequential skip brute force algorithm. However, the method is associated with network delays, communication and synchronization delays which would have affected the run time of the algorithm significantly.

In this paper, we are proving the hypothesis that using multicore CPUs to accelerate motif finding is better than accelerating in a grid network using the same skip-brute force algorithm and similar synthetic random data sets used by Faheem in [7].

## III. BACKGROUND

In this section we will give a brief description about the enhanced brute force algorithm and POSIX thread library used to implement the parallel version of the algorithm.

### A. Enhanced Brute Force Algorithm (Skip-Brute Force)

The skip brute force algorithm is an enhanced version of the brute force motif search algorithm used in a grid-based acceleration [6] and an FPGA based acceleration [7]. It is an exact algorithm for motif finding. As Brute Force algorithms use exhaustive enumeration, it first generates all the possible motifs (*l-mers*) of length *l*. All the *l-mers* are then compared with all the windows of each sequence (windows are all the substrings of length *l* of a DNA sequence starting at each position. If the length of the sequence is *n*, then the number of windows in the sequence is $(n - l) +1$).That is, the number of mismatches in nucleotides between each *l-mer* and the window is calculated. In other words the Hamming distance between the *l-mer* and the window is calculated. An *l-mer* is matched with a window if the number of mutations is less than or equal to the number of mutations (*d*) allowed. In the enhanced version, if any *l-mer* is not matched with at least one window, the algorithm skips checking the rest of the sequences and start comparison with the next *l-mer*. Therefore, all the unmatched *l-mers* are discarded eventually. The article [6] describes the pseudo code of this algorithm.

### B. POSIX Threads

A thread is a unit of execution within a program that can be executed independently of other codes. They can be used to implement parallelism in shared memory multicore architectures. POSIX Thread 1003.1c is the standard for UNIX systems. Libraries that have been developed using this standard are called POSIX threads. C and C++ languages have p_thread libraries.

## IV. DESIGN AND IMPLEMENTATION

In this section, we describe the approach taken to implement the parallel skip brute force program targeting a multicore CPU system.

### A. Thread Assignment Methodology

In the algorithm considered, there are a large number of comparisons between all these *l-mers* and the *windows* of each sequence. For example, for a pattern of length 10, $4^{10}$ *l-mers* should be compared with each *window*. These comparisons can be performed in parallel. Hence the approach for parallelization was to partition the number of *l-mers* that should be created by each thread. Number of *l-mers* that should be generated by each thread is calculated initially. The number of the first and the last *l-mer* each thread should generate are assigned as each thread data. There is no need of generating all the possible *l-mers* prior to the search. Basically each thread carries out the comparison process with only a portion of the *l-mers*. Hence the approach is data-level parallelism.

Figure 1 depicts the parallel approach. First two steps are performed sequentially. Inputs for the algorithm are the sequence file with the planted motifs, the length of the pattern to be searched, the number of mutations allowed and the number of threads. All the windows are first extracted from the sequence file. Then based on the number of threads used at a time, the number of *l-mers* each thread should generate is calculated. All threads get a copy of extracted windows and carry out skip-Brute Force search in parallel independently only with assigned *l-mers*. Each thread will output the identified motif.

### B. Implementation Details

First the synthetic datasets were generated according to the Fixed Motif Model (FM) described by the Pevzner and Sze [4]. Twenty random DNA sequences which are 600 nucleotides long and random motif consensuses of length from 11 to 15 nucleotides were created using RmotifGen software [13]. The probability of each nucleotide in the defined motif and the background sequence compositions were set to 25% for equal nucleotide frequency. Then 20 occurrences of each motif consensus of a specific length having d number of mutations were created randomly by selecting nucleotides randomly and planted in all sequences at random positions. Final result was 20 random DNA sequences of length 600 each having exactly one instance of the random motif. Data sets for the same (l, d) instances used in the grid-based implementation were created which are (11, 3), (12, 3), (13, 4), (14, 4) and (15, 4).

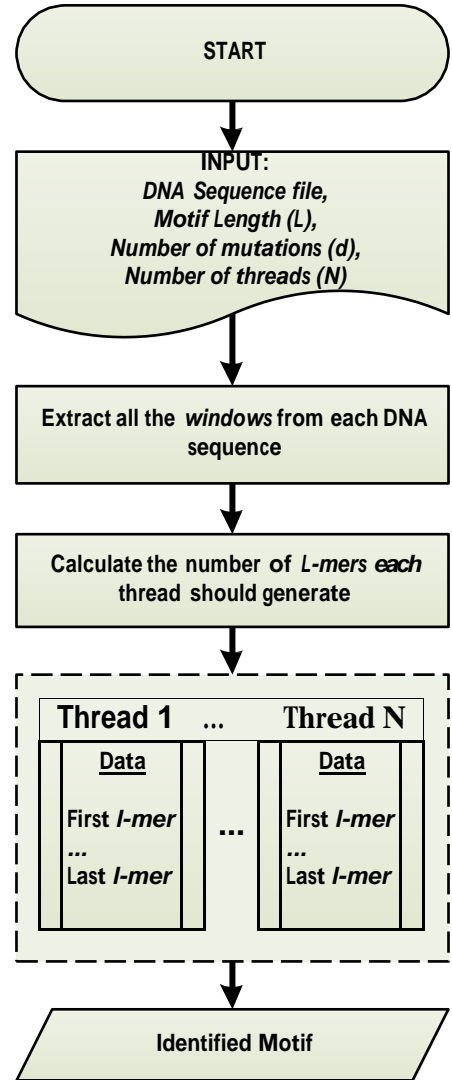

Fig.1.The thread based implementation of the enhanced brute force on Multicore CPUs

A sequential Skip-Brute Force algorithm was coded in C language. Parallel Skip-Brute Force algorithm was also coded in C language using POSIX thread library. Performances were measured by performing the same experiment several times and taking the average run time. Each experiment was repeated by increasing the number of threads used. This program was executed in different types of processors to check the effect of the processor configuration for the parallel program. All the tests were carried out on GNU/Linux platform.

## V. EXPERIMENTAL EVALUATION

Figure 2 shows the sequential skip-Brute Force run time implemented on an Intel Dual Core CPU. It shows exponential run time with the increase of the motif length. It takes more than 24 hours to solve the (15, 4) Challenge problem. Hence it is difficult to use for finding longer motifs.

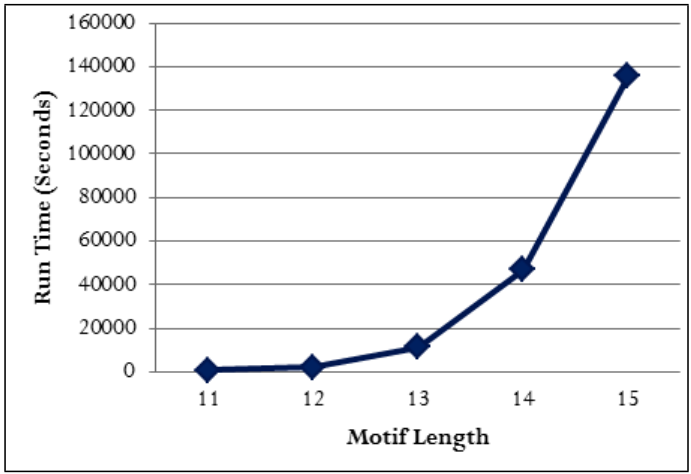

Fig. 2. The skip-BF run time that shows exponential run time with the increase of motif length

For the evaluation, the multi-threaded algorithm was tested in different types of processors to observe how the parallel algorithm utilizes the multi-threading supported by different multicore CPUs. Table I lists the different processors, on which the algorithm was tested. Processors P2-P4 consist of hyper-threading technology.

TABLE I
DIFFERENT PROCESSOR TYPES USED FOR TESTING THE PARALLEL SKIP BRUTE FORCE ALGORITHM

| Processor Number | Processor Description | Number of Cores | Optimal Number of threads |
|---|---|---|---|
| P1 | Intel Pentium Dual CPU T2330 (@2.20GHz) with 2GB RAM | 2 | 2 |
| P2 | Intel core i3 370 M(@3.30GHz) with 4GB RAM | 2 | 4 |
| P3 | Intel Core i5 (@ 3.30GHz) with 2GB RAM | 2 | 4 |
| P4 | 4 Intel Xeon CPU X7560 (@ 2.26 GHz) processors with a DDR3 RAM of 264GB. Each processor has 8 cores and can support 16 threads (hyper-threading) | 32 | 64 |

Figure 3 shows the speedup achieved by implementing the skip-BF algorithm in processors P1-P3 with the increase of number of threads. According to these graphs the speed up has been increased when the number of threads is increased up to the optimal number of threads. The behaviour of the graphs remains the same for the different motif sizes we tested. This shows that the program utilizes the parallelism in various processors by executing the comparison processes in parallel.

For example the implementation of this algorithm on a Dual core processor (P1) yielded a maximum average speedup of nearly 1.8x over the single thread implementation.

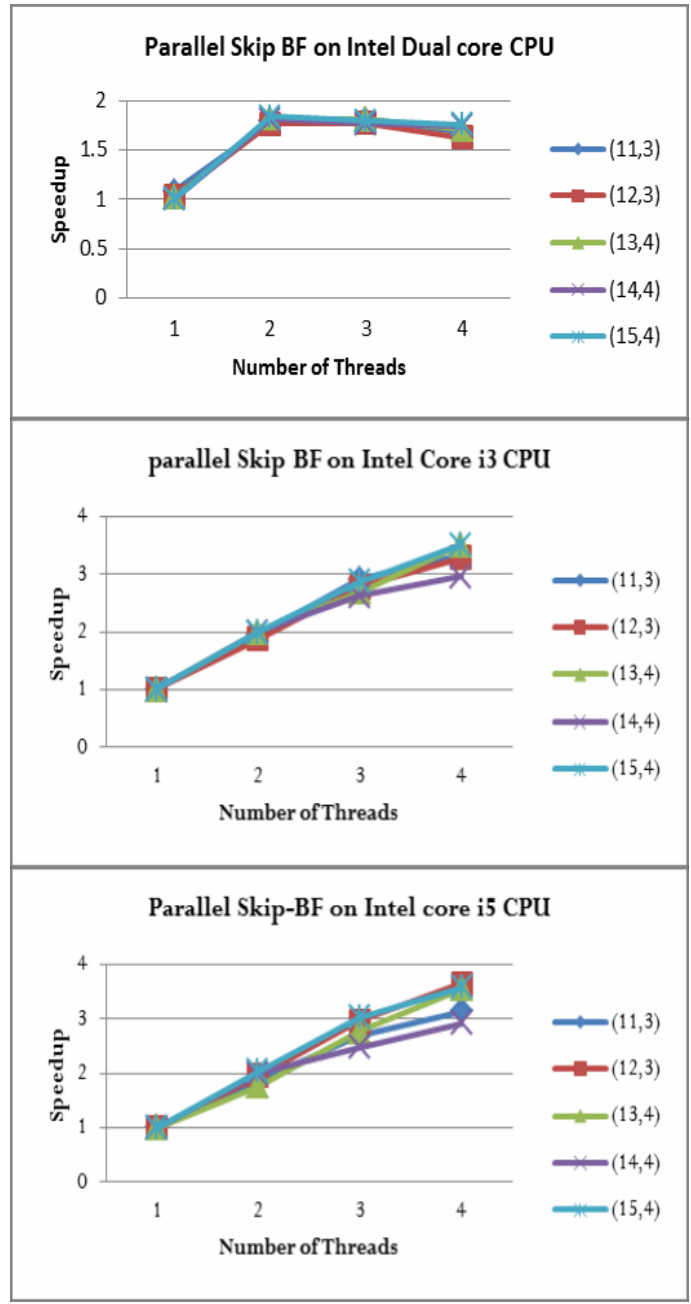

Fig. 3. The parallel skip-BF speedup on processors P1-P3

Since the dual core processor supports only for the maximum of two threads, the speed up saturates when the number of threads increases more than two. Both core i3 and core i5 processors support the maximum of four threads as stated in Table I. Hence the program achieves maximum speed up when four threads are used improving its performance.

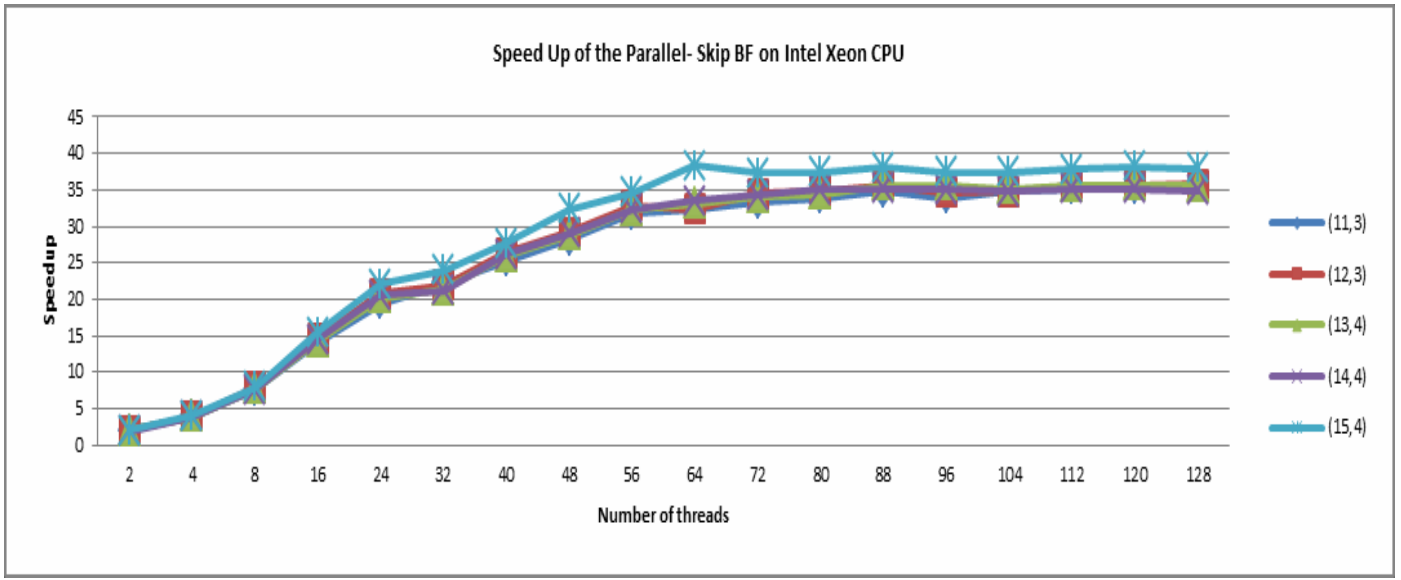

Fig. 4. The parallel skip-BF speedup on processor P4

Figure 4 depicts the speedup achieved by executing the parallel skip-BF on a 32 core processor over the single thread implementation. The speedup increases dramatically when the number of threads increases. The optimal number of threads supported by this processor is 64. Hence when number of threads increases more than 64 it does not increase the speed up.

TABLE II.
SPEEDUP COMPARISON BETWEEN MULTICORE AND GRID IMPLEMENTATIONS

| (l,d) values | Speedup | |
|---|---|---|
| | Grid Implementation on 32 nodes | Multicore Implementation with 32 cores |
| (11,3) | 13.258 | 32.390 |
| (12,3) | 19.418 | 32.308 |
| (13,3) | 22.133 | 33.152 |
| (14,4) | 23.156 | 33.537 |
| (15,4) | 23.608 | 38.348 |
| Average Speedup | ~21 | ~34 |

Additionally we have compared our method with a previous grid-based acceleration method [6]. Table II shows the comparison between speed ups achieved by the 32 cores in the multicore method and 32 grid nodes in the grid based architecture. The calculations for the grid implementation were performed using the timing results given for 24 and 96 worker nodes and the equations used in their calculations (assuming that the serial fraction is in between 24 and 96 implementations). The speedup for our 32 cores method was approximately 34x and the grid implementation was approximately 21x on a 32 grid nodes.

VI. CONCLUSION

In this paper a multicore CPU based method to accelerate motif finding in DNA sequences using Skip-Brute Force algorithm is presented. The parallel version of the algorithm was implemented and run in multicore architectures. According to the results obtained, we could conclude that the multicore machines can effectively be used for the motif finding problem. It was shown that the method achieved a significant speed up over the sequential version. Implementation on 32 core machine yielded an average maximum speed up of 34x which is much better than the average speed up of 21x if 32 grid nodes are used in grid based acceleration.

Our method was cost effective than other sophisticated parallel designs and time-efficient. Since special parallel hardware designs and complex programming were not required, it was easy to implement. Desktop computers with hyper-threading could be utilized easily. However the use of Multicore CPUs still depends on factors such as the type of the algorithm used, size of the data and the approach for the parallelization.

Future work of this research includes accelerating other exact and approximate motif finding algorithms such as Planted Motif Search (PMS), MEME, Projection, etc. using this method and acceleration using larger and real sequence data.